\documentclass[twocolumn,showpacs,preprintnumbers,superscriptaddress]{revtex4}
\usepackage{mathrsfs}
\usepackage{amssymb}
\usepackage{amsmath}
\usepackage{graphicx}
\usepackage{dcolumn}
\usepackage{bm}
\usepackage{graphicx}
\usepackage{float}
\usepackage[normalem]{ulem}
\usepackage{color}

\begin{document}
\title{Probing the Short Range Spin Dependent Interactions by Polarized $^{3}He$ Atom Beams}
\author{H.Yan}
\email[Corresponding author: ]{haiyan@umail.iu.edu}\affiliation{Institute of Nuclear Physics and Chemistry,China Academy of Engineering Physics, Mianyang, Sichuan, 621900,China }
\affiliation{Center for Exploration of Energy and Matter, Indiana University, Bloomington, IN 47408,USA}

\author{G.A.Sun}
\affiliation{Institute of Nuclear Physics and Chemistry,China Academy of Engineering Physics, Mianyang, Sichuan, 621900,China }

\author{J.Gong}
\affiliation{Institute of Nuclear Physics and Chemistry,China Academy of Engineering Physics, Mianyang, Sichuan, 621900,China }

\author{B.B.Pang}
\affiliation{Institute of Nuclear Physics and Chemistry,China Academy of Engineering Physics, Mianyang, Sichuan, 621900,China }

\author{Y.Wang}
\affiliation{Institute of Nuclear Physics and Chemistry,China Academy of Engineering Physics, Mianyang, Sichuan, 621900,China }

\author{Y.W.Yang}
\affiliation{Institute of Nuclear Physics and Chemistry,China Academy of Engineering Physics, Mianyang, Sichuan, 621900,China }

\author{J.Zhang}
\affiliation{Institute of Nuclear Physics and Chemistry,China Academy of Engineering Physics, Mianyang, Sichuan, 621900,China }

\author{Y.Zhang}
\affiliation{Institute of Nuclear Physics and Chemistry,China Academy of Engineering Physics, Mianyang, Sichuan, 621900,China }

\date{\today}
\begin{abstract}
Experiments using polarized $^{3}He$ atom beams to search for short range spin dependent forces are proposed. High intensity, high polarization, small beam size
$^{3}He$ atom beams have been successfully produced and used in surface science researches. By incorporating background reduction designs
as combination shielding by $\mu$-metal and superconductor and double beam paths, the precision of spin rotation angle per unit length could be improved by
a factor of $\sim10^{4}$.
By this precision, in combination with using a high density and low magnetic susceptibility sample source mass,and reversing one beam path if necessary,
sensitivities on three different types of spin dependent interactions
 could be possibly improved by as much  as $\sim10^{2}$ to $\sim10^{8}$ over the current experiments at the millimeter range.
\end{abstract}
\maketitle
\section{Introduction}
 New physics beyond the standard model is possible. Various new particles as axions, familons and majorons,etc\cite{PDG12} with masses were theoretically introduced.
 New macroscopic interactions meditated by WISPs (weakly-interacting sub-eV particles) could exist. These new possible forces have
ranges from nanometers to meters.The fact that the dark energy density of order (1 meV)${^4}$  corresponds to a length scale of $~100$ $\mu$m  also encourages
 searches for new phenomena around this scale~\cite{ADE09}. 
In Ref.\cite{DOB06}, interactions between non-relativistic  fermions assuming only
rotational invariance can be classified into 16 different operator structures involving the spin and momenta of the particles.
For all these sixteen interactions, only one interaction does not require either of the two particles to be spins polarized; six interactions require at least
one particle to be spin polarized and the remaining nine require both particles  to be spin polarized.\\

Experimental constraints on possible new interactions of mesoscopic range which depend on the spin of one or both of the particles are much less stringent than those for spin-independent interactions. This is not surprising since macroscopic objects with large nuclear or electron polarization are not easy to arrange outside an environment that includes large magnetic fields, which can produce large systematic effects in delicate experiments. On the other hand the addition of the spin degree of freedom opens up a large variety of possible new interactions to search for which might have escaped detection to date.\\

Among the six type interactions which only one particle needs to be spin polarized, the scalar-pseudoscalar interaction $V_{SP}(r)$($V_{9,10}$ in Ref.\cite{DOB06}'s notation ) originated from  the coupling $\mathcal{L}_{\phi}=\bar{\psi}(g_{s}+ig_{p}\gamma_{5})\psi\phi$\cite{MOO84,PIE11},or the monopole dipole interaction
has begun to attract more scientific attention recently. The interaction between the polarized spin $1/2$ fermion of mass $m$ and another unpolarized nucleon can be expressed as:
\begin{equation}\label{eqnSP}
V_{SP}(r)=\frac{\hbar^{2}g_{S}g_{P}}{8\pi m}(\frac{1}{\lambda r}+\frac{1}{r^{2}})\exp{(-r/\lambda)}\vec{\sigma}\cdot\hat{r}
\end{equation}
where $\lambda=\hbar/m_{\phi}c$ is the interaction range, $m_{\phi}$ is the mass of the new scalar boson,$\vec{s}=\hbar\vec{\sigma}/2$ is the spin of the
polarized particle and $r$ is the distance between the two interacting particles.
While for the vector-axial-vector interaction $V_{VA}(r)$($V_{12,13}$ in Ref.\cite{DOB06}'s notation) originated from the coupling $\mathcal{L}_{X}=\bar{\psi}(g_{V}\gamma^{\mu}+g_{A}\gamma^{\mu}\gamma_{5})\psi X_{\mu}$,
this parity violating interaction has the form:
  \begin{equation}\label{eqnVA}
  V_{VA}(r)=\frac{\hbar g_{V}g_{A}}{2\pi}\frac{\exp{(-r/\lambda)}}{r}\vec{\sigma}\cdot\vec{v}
  \end{equation}
where $\vec{v}$ is the relative velocity between the probe particle and source particle,$\lambda=\hbar/m_{X}c$ is the interaction range, $m_{X}$ is the mass of the new vector boson.
$V_{VA}(r)$ is the Yukawa potential
times the $\vec{\sigma}\cdot\vec{v}$ factor,which makes this interaction quite interesting. Another interaction requiring only one particle to be spin polarized is
the axial-axial interaction $V_{AA}(r)$($V_{4,5}$ in Ref.\cite{DOB06}'s notation), which is also originated from the $\mathcal{L}_{X}$ coupling, can be written as:
\begin{equation}\label{eqnAA}
V_{AA}(r)=\frac{\hbar^{2}g_{A}^{2}}{16\pi mc}(\frac{1}{\lambda r}+\frac{1}{r^{2}}) {\exp{(-r/\lambda)}}\vec{\sigma}\cdot(\vec{v}\times\hat{r})
\end{equation}
All these interactions are in the form of $\vec{s}\cdot\vec{B'}$ where $\vec{B'}$ can be viewed as a pseudo magnetic field\cite{PIE11}.
For an unpolarized source mass as a plane plate of thickness $d$ and surface normal vector $\hat{y}$,
as in Ref.\cite{ZIM10,PIE11}, for a spin polarized probe particle moving with velocity $\vec{v}$,   the corresponding pseudo-magnetic fields due to these three interactions can be derived as:
\begin{eqnarray}\label{eqnPB}
\vec{B}_{SP}=\frac{1}{\gamma}\frac{\hbar g_{S}g_{P}}{2m}\rho_{N}\lambda e^{-\frac{\Delta y}{\lambda}}[1-e^{-\frac{d}{\lambda}}]\hat{y}\\
\vec{B}_{VA}=\frac{2}{\gamma}g_{V}g_{A}\rho_{N}\lambda^{2} e^{-\frac{\Delta y}{\lambda}}[1-e^{-\frac{d}{\lambda}}]\vec{v}\\
\vec{B}_{AA}=\frac{1}{\gamma}\frac{g_{A}^{2}}{4}\rho_{N}\frac{\hbar}{mc}\lambda e^{-\frac{\Delta y}{\lambda}}[1-e^{-\frac{d}{\lambda}}]\vec{v}\times\hat{y}
\end{eqnarray}
where $\Delta y>0$ is the distance from the probe particle to the sample surface,$\rho_{N}$ is the nucleon number density
of the sample,$\gamma$ the gyromagnetic ratio of the probing particle. \\

Various experiments have been performed to search for such interactions. In Ref.\cite{PET10,FU11,ZHE12,BUL12,CHU13,TUL13}, polarized noble gases as
$^{3}He$,$^{129}Xe$ and $^{131}Xe$ have been used to probe the $V_{SP}(r)$ force. In Ref.\cite{PIE11}, a table-top neutron
Ramsey apparatus is suggested to search for the  $V_{VA}(r)$  and the $V_{AA}(r)$ interaction. The Ramsey's separated oscillating field technique has a very high sensitivity
since it converts the quantum mechanical phase shift into a frequency difference.
 The proposed experiment to probe the $V_{AA}(r)$ was performed and reported in Ref.\cite{PIE12}, the results provide the most stringent constraint on $g_{A}g_{A}$ in the millimeter range. \\

  Very recently, a constraint on $g_{V}g_{A}$ by probing $V_{VA}(r)$ through measuring neutron spin rotation in liquid $^{4}He$ was reported in Ref.\cite{YAN13}. Though
 the neutron spin rotation experiment was originally designed to detect the parity violating weak NN interaction, it turns out to be so far the  most sensitive method for probing
 the spin-velocity interaction(Eqn.(\ref{eqnVA})) in short ranges. The derived constraint is more than $\sim10^{7}$ times stringent than the current existing laboratory constraints in micrometer
 to meter ranges. However,it would be hard to further improve the sensitivity of this experiment method. On one hand, the current sensitivity is mainly limited by statistics,i.e.,the neutron
 counts,for which the flux is not very easy to increase by orders and the neutron beam integration time was already as long as $\sim 1$ year. On the other hand, even the sensitivity can be
 improved, the parity violation background from the standard model will be a problem. The expected size of the parity-odd rotation angle in liquid Helium due to
 the standard model is about $10^{-6}$ to $10^{-7}$ rad/m\cite{STO74}, and
 the performed measurement has already achieved a precision in this regime. If the precision is improved and a nonzero spin rotation is observed, it is impossible to tell whether it is from the new interaction or the standard model since the later can not be precisely calculated yet.\\
\section{The Proposed Experiment Scheme}
\begin{figure*}
\begin{center}
\includegraphics[scale=0.6,angle=0]{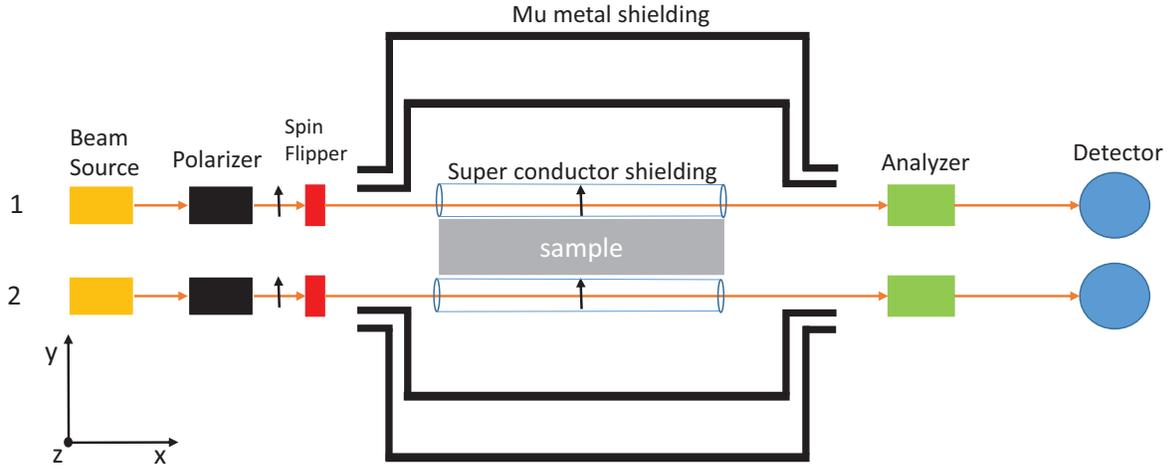}
\caption{Color online, schematic drawing of the proposed experiment set up,top view .}
\label{fig:apparatus}
\end{center}
\end{figure*}
\begin{figure}[h]
\centering
 \includegraphics[scale=1.2, angle=0]{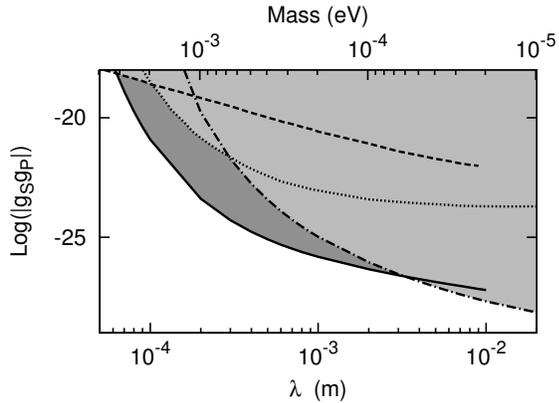}
 \caption{\small{Expected sensitivity(solid line) of the proposed experiment for the Scalar-Pseudoscalar interaction. The light grey area is the excluded area
 by present experiments. The dashed line is the result of \cite{ASZ04}, the doted line is the result of \cite{BUL12}, the dash-dotted line is the result of
  \cite{TUL13}. }}
 \label{fig.gsgp}
\end{figure}
\begin{figure}[h]
\centering
 \includegraphics[scale=1.2, angle=0]{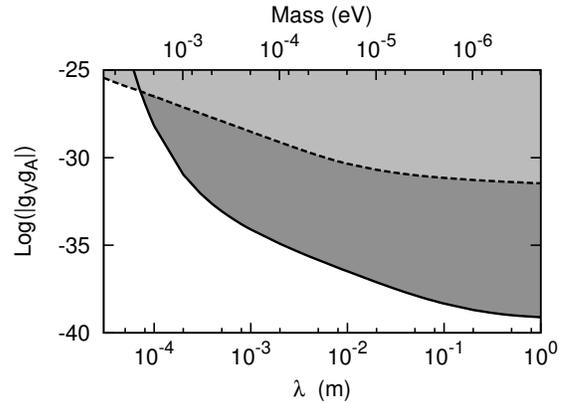}
 \caption{\small{Expected sensitivity(solid line) of the proposed experiment for the Vector-Axial interaction. The light grey area is the excluded area
 by present experiments. The dashed line is the result of \cite{YAN13}. }}
 \label{fig.gvga}
\end{figure}
\begin{figure}[h]
\centering
 \includegraphics[scale=1.2, angle=0]{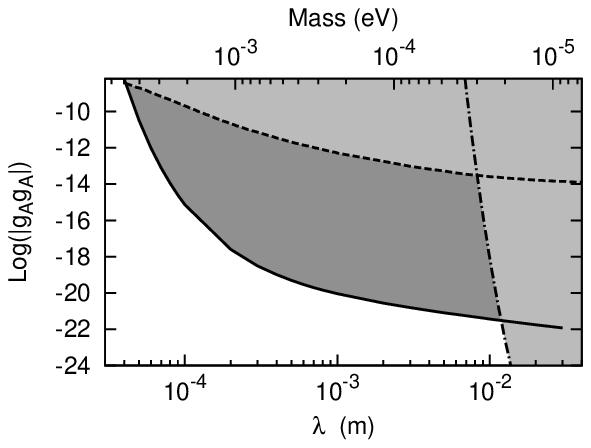}
 \caption{\small{Expected sensitivity(solid line) of the proposed experiment for the Axial-Axial interaction. The light grey area is the excluded area
 by present experiments. The dashed line is the result of \cite{PIE12}, the dash-dotted line is the result of \cite{VAS12}.}}
 \label{fig.gaga}
\end{figure}
For the above mentioned new interaction probing methods, when using polarized noble gases, large quantities of probing particles
can be obtained but the polarized gases has to be stored in certain magnetic fields to keep their polarizations, thus
systematics due to the nonzero background fields are unavoidable. The polarized nobel gases are usually sealed in glass cells and the glass wall will limit the probe to source distance. Besides, it would be technically difficult to  realize a large relative velocity between the source mass and the probe of polarized nobel gases sealed in glass cells. Thus it would not be easy to detect the velocity dependent new interactions using polarized noble gases. 
When using neutron beams, the field background can be
greatly reduced and the probe to source distance can actually be zero  as in Ref.\cite{YAN13,SNO11}  but the quantity of the probing particles is limited by the available neutron flux.
In order to further improve the sensitivity of detecting the spin dependent short range interactions, we propose an experiment
using nuclear spin polarized $^{3}He$ atom beams. Though in principle other spin-$1/2$ species as $^{129}Xe$ might also work,
the polarized $^{3}He$  beam technique is more convenient since it has been well developed and applied to study the surface dynamics
in condensed matter physics for many years\cite{DEK95,JAR01}. Ref.\cite{JAR09} is a very nice review for the recent developments of the so called polarized $^{3}He$
spin-echo technique. The schematic drawing of the proposed experiment is shown as Fig.1. The $^{3}He$ beam is firstly
produced by the standard atomic beam method\cite{MIL88} of expanding compressed $^{3}He$ gas through a fine nozzle into vacuum. The speed  of the beam can be controlled by adjusting the nozzle temperature. Then the beam is polarized by a beam polarizer which is made of hexapole magnets\cite{JAR01}. High intensity ($1.5\times10^{14}atoms/s$ reported in Ref.\cite{FOU05}), small size($2mm$ beam diameter at target according to Ref.\cite{FOU05}) and high polarization (more than $90\%$ reported in Ref.\cite{DEK95}) $^{3}He$ beams can be produced. The
polarized $^{3}He$ beam will then fly over the surface of the high density sample as a lead plate. The beam polarization will be rotated by the new spin dependent interactions if exist.\\

To reduce the background, the sample is firstly shielded with multiple cylindrical layers of high permeability materials as the $\mu$-metal. In this way, the background field could
 initially be reduced to $\sim10^{-9}T$\cite{KOR07}. Then the sample is further shielded by several superconducting layers as thin lead foils. The residual field could
be reduced to be less than $10^{-11}T$\cite{PAV98}. After passing the sample area, the beam goes through an analyzer which is  another series of hexpole magnets\cite{DWO04}. Only the  right polarization state atoms are focused and can go through the analyzer to reach the helium detector.
To further reduce the background, a double beam design is applied. As
shown by Fig.1, another polarized beam produced by the same way will fly over the other sample surface, then it will be analyzed and detected similarly. The new spin dependent interaction
 signal can be extracted from the rotation difference between the two beam spins. In Ref.\cite{SNO11,YAN13}, the uniformity of the residual background field reaches $10^{-4}$ level
  for a $5cm\times5cm$ beam size. In this work, a higher uniformity is expected since the beam size is much smaller($2mm$ beam diameter). Thus from the difference, at most, a $\sim10^{-15}T$ background is estimated and it is considered to be the main systematic for the proposed experiment. There are no systematics from the standard model since now the probe particle
   does not contact with the sample directly.\\

For different spin dependent interactions, different polarization and beam path arrangements can be made to detect the specified interaction. In more detail, for the interaction $V_{SP}(r)$,
the pseudo-magnetic field for beam 1 of Fig.1 is along $+\hat{y}$ direction while for beam 2 $-\hat{y}$ direction. For $V_{VA}(r)$, the pseudo-magnetic field direction is along the beam moving direction either for beam 1 or beam 2. For $V_{AA}(r)$, the pseudo-magnetic field is along $+\hat{z}$ direction for beam 1 and $-\hat{z}$ for beam 2. To detect $V_{SP}$, both beam 1 and  beam 2 can be set to be polarized along $+\hat{z}$ or $-\hat{z}$ direction. The difference of the spin rotation angles between the two beams will cancel the common background field
effect and only leave the pseudo-magnetic field effect since it induces opposite rotation angle for each beam. Since the beams are polarized along $\hat{z}$, the
spin rotation difference will not be sensitive to $V_{AA}(r)$ which is along the $\pm\hat{z}$ or $V_{VA}(r)$ which will rotate the two beam polarizations along $\hat{v}=+\hat{x}$ by the
 same amounts. Similarly, $V_{AA}(r)$ can be detected by setting the both beam polarizations along $+\hat{y}$ or $-\hat{y}$ direction. To detect $V_{VA}(r)$, one of the beam path could be flipped thus the relative velocity between the probe beam and the source sample is reversed. If the beam polarization is along $\hat{y}$ direction, the reversed beam setup will be only sensitive to $V_{VA}(r)$. This beam path reversing feature
 is possible for the atomic beam techniques since all the components are compact enough while it is not easy to be realized for the neutron beams without losing intensity significantly.\\
\section{The Sensitivity of the Proposed Experiment}
 By carefully arranging more polarization combinations, and in combination of moving source mass in and out if possible,
  the background induced systematics might be further reduced. When the probe beams reach the detectors, the spin rotation angle $\phi$ can be obtained from the
  beam counts\cite{SNO11}:
 \begin{equation}
 \phi\approx\sin{\phi}=\frac{N_{+}-N_{-}}{N_{+}+N_{-}}\frac{1}{AP}
 \end{equation}
 where $N_{+}$ and $N_{-}$ are the counts for spin up and down states measured by the helium atom detectors,$A$ and $P$ are the Analyzing power and Polarizing power of the beam polarizer and analyzer respectively. Assuming the beam integration time is $\sim100$ days, with the known analyzing power, polarizing power of the magnets, and the best known detector efficiency\cite{JAR09}, for a meter long lead sample with thickness of $10cm$, the sensitivity of the spin rotation angle per unit length for probe particle speed of $1000m.s^{-1}$ is found to be
 \begin{equation}
 \frac{d\phi}{dL}\sim\pm3.5(stat)\pm2.0(sys)\times10^{-10}m^{-1}
 \end{equation}\\

Based on this sensitivity, assume the beam to sample distance is $\sim1mm$, the constraints on $V_{SP}$, $V_{VA}$ and $V_{AA}$ are plotted as FIG.2,3 and 4 respectively. For the constraint on $g_{S}g_{P}$, the proposed experiment could improve as much as $\sim10^{2}$ in ranges of $100\mu m$ to $1mm$. For $g_{V}g_{A}$, the new experiment can improve sensitivity by as much as $\sim10^{7}$  in ranges of $100\mu m$ to $1m$. For $g_{A}g_{A}$, the constraint will be improved by more than $\sim10^{8}$ in the ranges of
$100\mu m$ to $1cm$.
\section{conclusion and discussion}
In summary, we proposed an experiment scheme for which the sensitivity of detecting the single spin dependent interactions  could be improved by many orders at
mesoscopic ranges. This new scheme combines the high intensity, easy beam maneuverability of the atom beam techniques with the improved background reduction design of
the neutron spin rotation experiment. The sensitivity of spin rotation angle per unit length can be improved by $10^{4}$ times.
Using the high intensity sample as a thick lead plate, high sensitivities on probing single spin dependent interactions can be achieved.
For the scalar-pseudo scalar interaction, the improvement mainly is in range $10^{-4}\sim10^{-3}m$, as many as $\sim 2$ orders of sensitivity can be improved.
For the vector-axial interaction, as much as $\sim 7$ orders sensitivity can be improved for ranges below $\sim 1m$. For the axial-axial spin dependent interaction, the sensitivity could be improved by more than $\sim 8$ orders.
 The sensitivity improvement on $g_{S}g_{P}$ is not as impressive as for $g_{V}g_{A}$ and $g_{A}g_{A}$.
The reason is that the $\sim10^{-10}rad.m^{-1}$ spin rotation per unit length precision is actually one order lower than the frequency shift precision achieved in Ref.\cite{TUL13} and the improvement is mainly from a shorter probe to source distance. The method proposed here is more advantageous for detecting $g_{V}g_{A}$ and $g_{A}g_{A}$.\\

It should be pointed out that several estimations made here are considered to be conservative. For example, $10^{-11}T$ background residual field was assumed while $\sim10^{-12}T$ background level \cite{CAB73,CAB88} had been achieved in 1970's for a meter long cylindrical structure with a $20cm$ diameter and opening. While the diameter of the innermost meter long
shielding for the proposed experiment would be less than $\sim1$cm, since the beam size is only $\sim2$mm,
 thus better shielding\cite{BUD13} and less background field would be expected. The $10^{-4}$ uniformity was assumed according to the level of neutron spin rotation apparatus which is for a much larger beam size. The statistical error is derived on assumption of $7\times10^{-3}$ helium detector efficiency which still has many space to improve. Several key parameters are borrowed from the Helium spin echo techniques. It will not be surprising that the sensitivity can be further improved for a dedicated experiment searching for the new spin dependent interactions.\\

We acknowledge support from the National Natural Science Foundation of China, under grant 11105128, 91126001 and 51231002.  H.Y. is supported by the US  NSF grant PHY-1207656. We thank Dr.  A.P.Jardine for helpful discussions and useful references. We thank Dr. M.W.Snow for inspiring comments.  H.Y. thanks Professor J.Long for support. We would like to thank the anonymous referee for the valuable comments and detailed corrections, which we found very constructive and helpful to improve our manuscript.

\end{document}